\documentclass[twocolumn,tighten,preprint2]{aastex62}
\usepackage{apjfonts}
\usepackage{color}
\usepackage{amsmath, mathrsfs}

\shorttitle{Long-term Optical Variations of Ark 120} 
\shortauthors{Li et al.}
\begin{document}

\title{A Possible $\sim$20 yr Periodicity in Long-term Optical Photometric and Spectral Variations of the Nearby Radio-Quiet \\
Active Galactic Nucleus Ark 120}

\author[0000-0001-5841-9179]{Yan-Rong Li}
\affiliation{Key Laboratory for Particle Astrophysics, Institute of High 
Energy Physics, Chinese Academy of Sciences, 19B Yuquan Road, 
Beijing 100049, China}

\author[0000-0001-7617-4232]{Jian-Min Wang}
\affiliation{Key Laboratory for Particle Astrophysics, Institute of High 
Energy Physics, Chinese Academy of Sciences, 19B Yuquan Road, 
Beijing 100049, China}
\affiliation{National Astronomical Observatories of China, Chinese 
Academy of Sciences, 20A Datun Road, Beijing 100012, China}
\affiliation{School of Astronomy and Space Science, University of Chinese Academy of Sciences, 
19A Yuquan Road, Beijing 100049, China}

\author{Zhi-Xiang Zhang}
\author{Kai Wang}
\author{Ying-Ke Huang}
\affiliation{Key Laboratory for Particle Astrophysics, Institute of High 
Energy Physics, Chinese Academy of Sciences, 19B Yuquan Road, 
Beijing 100049, China}

\author[0000-0003-2112-171X]{Kai-Xing Lu}
\affiliation{Yunnan Observatories, Chinese Academy of Sciences, Kunming 650011, China}

\author[0000-0001-8492-6369]{Chen Hu}
\author[0000-0002-5830-3544]{Pu Du}
\affiliation{Key Laboratory for Particle Astrophysics, Institute of High 
Energy Physics, Chinese Academy of Sciences, 19B Yuquan Road, 
Beijing 100049, China}

\author[0000-0002-0465-8112]{Edi Bon}
\affiliation{Astronomical Observatory, Belgrade, Serbia}

\author[0000-0001-6947-5846]{Luis~C.~Ho}
\affiliation{ Kavli Institute for Astronomy and Astrophysics, Peking University, 
Beijing 100871, China}
\affiliation{Department of Astronomy, School of Physics, Peking University, 
Beijing 100871, China}

\author{Jin-Ming Bai}
\affiliation{Yunnan Observatories, Chinese Academy of Sciences, Kunming 650011, China}

\author[0000-0002-2121-8960]{Wei-Hao Bian}
\affiliation{Physics Department, Nanjing Normal University, Nanjing 210097, China}

\author[0000-0002-7330-4756]{Ye-Fei Yuan}
\affiliation{Department of Astronomy, University of Science and Technology of China, Hefei 230026, China}

\author[0000-0003-2662-0526]{Hartmut Winkler}
\affiliation{Department of Physics, University of Johannesburg, PO Box 524, 2006 Auckland Park, Johannesburg, South Africa}

\author{Eduard K. Denissyuk}
\author{Rashit R. Valiullin}
\affiliation{V.G. Fesenkov Astrophysical Institute, Almaty, Republic of Kazakhstan}

\author[0000-0002-3462-4888]{Nata\v{s}a Bon}
\affiliation{Astronomical Observatory, Belgrade, Serbia}

\author[0000-0003-2398-7664]{Luka {\v C.} Popovi{\'c}}
\affiliation{Astronomical Observatory, Belgrade, Serbia}
\affiliation{Isaac Newton Institute of Chile, Yugoslavia Branch Belgrade, Serbia}

\email{liyanrong@mail.ihep.ac.cn, wangjm@mail.ihep.ac.cn}

\begin{abstract}
We study the long-term variability in the optical monitoring database of Ark~120, 
a nearby radio-quiet active galactic nucleus (AGN) at a distance of 143 Mpc ($z=0.03271$). We compiled the historical 
archival photometric and spectroscopic data since 1974 and conducted 
a new two-year monitoring campaign in 2015$-$2017, resulting in a total temporal baseline over 
four decades. The long-term variations in the optical continuum exhibit a wave-like pattern and the H$\beta$ integrated 
flux series varies with a similar behavior. The broad H$\beta$ profiles have asymmetric double peaks, which change strongly with time and 
tend to merge into a single peak during some epochs.
The period in the optical continuum determined from various period-search methods is about $20$ yr and the estimated false alarm probability 
with null hypothesis simulations is about $1\times10^{-3}$.
The overall variations of the broad H$\beta$ profiles also follow the same period. However,
the present database only covers two cycles of the suggested period, which strongly encourages continued monitoring to track more cycles 
and confirm the periodicity. Nevertheless, in light of the possible periodicity and the complicated H$\beta$ profile, 
Ark~120 is one candidate of the nearest radio-quiet AGNs with possible periodic variability, and it is thereby a potential candidate 
host for a sub-parsec supermassive black hole binary. %
\end{abstract}

\section{Introduction}

\begin{deluxetable*}{lccccccccc}
\tablecolumns{6}
\tablewidth{1.0\textwidth}
\tabletypesize{\footnotesize}
\tablecaption{Data Sets for the light curves of Ark~120.}
 \tablehead{
 \colhead{Data set}   &
 \colhead{~~~References~~~}      &
 \multicolumn{2}{c}{Observation Period} &
 \colhead{Number of}  &
 \colhead{Aperture}  &
 \colhead{$F(\text{[\ion{O}{3}]})$\tablenotemark{$a$} } &
 \colhead{$\varphi$} &
 \colhead{$G$}\\\cline{3-4}
 \colhead{}   &
 \colhead{}   &
 \colhead{(JD-2,400,000)}   &
 \colhead{(Year)}   &
 \colhead{Obs.}   &
 \colhead{(arcsec)}   &
 \colhead{($10^{-13}{\rm~erg~s^{-1}~cm^{-2}}$)}&
 \colhead{}  &
 \colhead{($10^{-15}{\rm~erg~s^{-1}~cm^{-2}~\text{\AA}^{-1}}$)} 
 }
\startdata
\sidehead{Spectroscopy}\hline
D99b        & 1   & 42,392-47,944 & 1974-1990   & 63  &  $2.0\times7.0$ &   1.35    & $1.008\pm0.017$ & $\phm{+}2.05\pm0.25$\\
P89         & 2   & 44,168-47,414 & 1979-1988   & 74  &  $7.0$          &   0.925   & $1.005\pm0.021$ & $\phm{+}1.39\pm0.32$\\
P98         & 3   & 47,524-50,388 & 1988-1996   & 141 &  $5.0\times7.6$ &   0.91    & 1  & 0 \\
W96         & 4   & 48,619-48,834 & 1991-1992   & 20  &  $5.0\times10.0$&   0.70    & $1.078\pm0.005$ & $\phm{+}0.46\pm0.06$\\
D08         & 5   & 48,629-53,445 & 1992-2005   & 88  &  $3.0\times11.0$&   1.13    & $1.037\pm0.003$ & $\phm{+}0.09\pm0.03$\\
L           & 6   & 57,295-57,839 & 2015-2017   & 100 &  $2.5\times8.5$  &   0.96    & \nodata       & \nodata      \\\hline
\sidehead{Photometry\tablenotemark{$b$}}\hline
M79\tablenotemark{c}         & 7    & 25,622-33,446 & 1935-1950  & 52  &  \nodata        &   \nodata  & \nodata        & \nodata\\
P83         & 8    & 42,392-44,226 & 1974-1979  & 36  &  $15.0$         &   \nodata  & $0.687\pm0.087$ & $\phm{+}0.45\pm0.95$ \\
P89         & 2    & 43,426-47,414 & 1977-1988  & 74  &  $7.0$          &   \nodata   & $0.607\pm0.297$ & $\phm{+}5.04\pm0.34$ \\
D99a1       & 9    & 43,963-50,818 & 1979-1998  & 25  &  $14.3$         &   \nodata   & $1.132\pm0.070$ & $-5.04\pm0.90$\\
D99a2       & 9    & 43,079-51,186 & 1976-1999  & 50  &  $27.5$         &   \nodata   & $1.126\pm0.064$ & $-8.07\pm1.03$ \\
W92/97\tablenotemark{$d$}         & 10    & 47,534-52,282 & 1989-2002  & 37  &  $20.0$         &   \nodata   & $0.719\pm0.017$ & $-0.79\pm0.28$ \\
R12         & 11   & 52,207-55,644 & 2001-2011  & 54  &  \nodata        &   \nodata  & $0.787\pm0.058$ & $-0.13\pm0.85$  \\
K14         & 12   & 52,237-54,319 & 2001-2007  & 67  &  $8.3$          &   \nodata  & $0.903\pm0.030$ & $-0.34\pm0.40$ \\
C\tablenotemark{$e$}       & 13   & 53,709-56,580 & 2005-2013  & 73  &  \nodata        &   \nodata  & $0.358\pm0.015$ & $\phm{+}6.51\pm0.22$ \\
H\tablenotemark{$f$}           & 14         &      55,122-55,268   & 2009-2010  & 96  &  7.5            &   \nodata  &  1    &  $-7.86\pm0.05$\\
A           & 15   & 55,949-58,345 & 2012-2018  & 476 &  \nodata        &   \nodata  & $0.808\pm0.005$ & $-2.97\pm0.07$
\enddata 
\tablerefs{ (1) \cite{Doroshenko1999b}; (2) \cite{Peterson1989};     (3) \cite{Peterson1998};
            (4) \cite{Winge1996};       (5) \cite{Doroshenko2008};  (6) this work;      (7) \cite{Miller1979}
            (8) \cite{Peterson1983};    (9) \cite{Doroshenko1999a};  (10) \cite{Winkler1992} and \cite{Winkler1997};
            (11) \cite{Roberts2012};     (12) \cite{Koshida2014};    (13) the Catalina Real-Time Survey database (\citealt{Drake2009});  
            (14) \cite{Haas2011};        (15) the All-Sky Automated Survey for Supernovae database (\citealt{Shappee2014, Kochanek2017}).
}
\tablenotetext{a}{The flux of [\ion{O}{3}]~$\lambda5007$ used for absolute calibration of spectra in the dataset.}
\tablenotetext{b}{The photometric data of all the datasets are at $V$-band, except for dataset M79 at $B$-band.}
\tablenotetext{c}{This dataset has no temporal overlap with the other datasets so that it is impossible to do 
intercalibration. We simply adjust the fluxes to make the mean and standard deviation 
to be at the same scales as those of the other datasets. We do not use this dataset for periodicity analysis.}
\tablenotetext{d}{This dataset includes 15 additional observations (50,863-52,282) that have not been public.}
\tablenotetext{e}{There is a large flux difference with respect to the other datasets. We offset the fluxes by 
$-20.0\times10^{-15}{\rm~erg~s^{-1}~cm^{-2}~\text{\AA}^{-1}}$ before performing intercalibration. }
\tablenotetext{f}{This dataset has a poor temporal overlap with the other datasets and we simply fix $\varphi=1$. }
\label{tab_lc}
\end{deluxetable*}

Periodic brightness variability in long-term monitoring of active galactic nuclei (AGNs) is widely used to 
search for supermassive black hole (SMBH) binary candidates in modern time-domain surveys (e.g., 
\citealt{Graham2015a, Graham2015b, Liu2015, Liu2016, Charisi2016, Zheng2016}). Although there are alternative %
explanations, such a periodicity is generally believed to  reveal the binary's orbital motion, which %
modulates accretion processes into the black holes, giving rise to periodic variations in disk emissions.
To date, systematic searches from large surveys have yielded more than one hundred quasar and AGN 
candidates with periodic variability, distributed over a wide redshift range up to $z\sim3$ (\citealt{Graham2015b, Charisi2016, Liu2016}). 
Due to the limited temporal baselines ($\lesssim$10 yr), these detected rest-frame periods are confined to short 
timescales of several years or hundreds of days. There are also a number of quasars and nearby AGNs that were reported
individually to exhibit periodicity based on long-term databases, including (but not limited to) OJ~287 ($z=0.31$; \citealt{Valtonen2008}), %
NGC 4151 ($z=0.003$; \citealt{Guo2006, Bon2012}), and NGC 5548 ($z=0.017$; \citealt{Bon2016,Li2016}). OJ 287 is long-known to have 
strong radio emission, while NGC 4151 and NGC 5548 have relatively weak radio emission (\citealt{Ho2002}). %

\begin{figure*}[t!]
\centering
\includegraphics[width=0.495\textwidth]{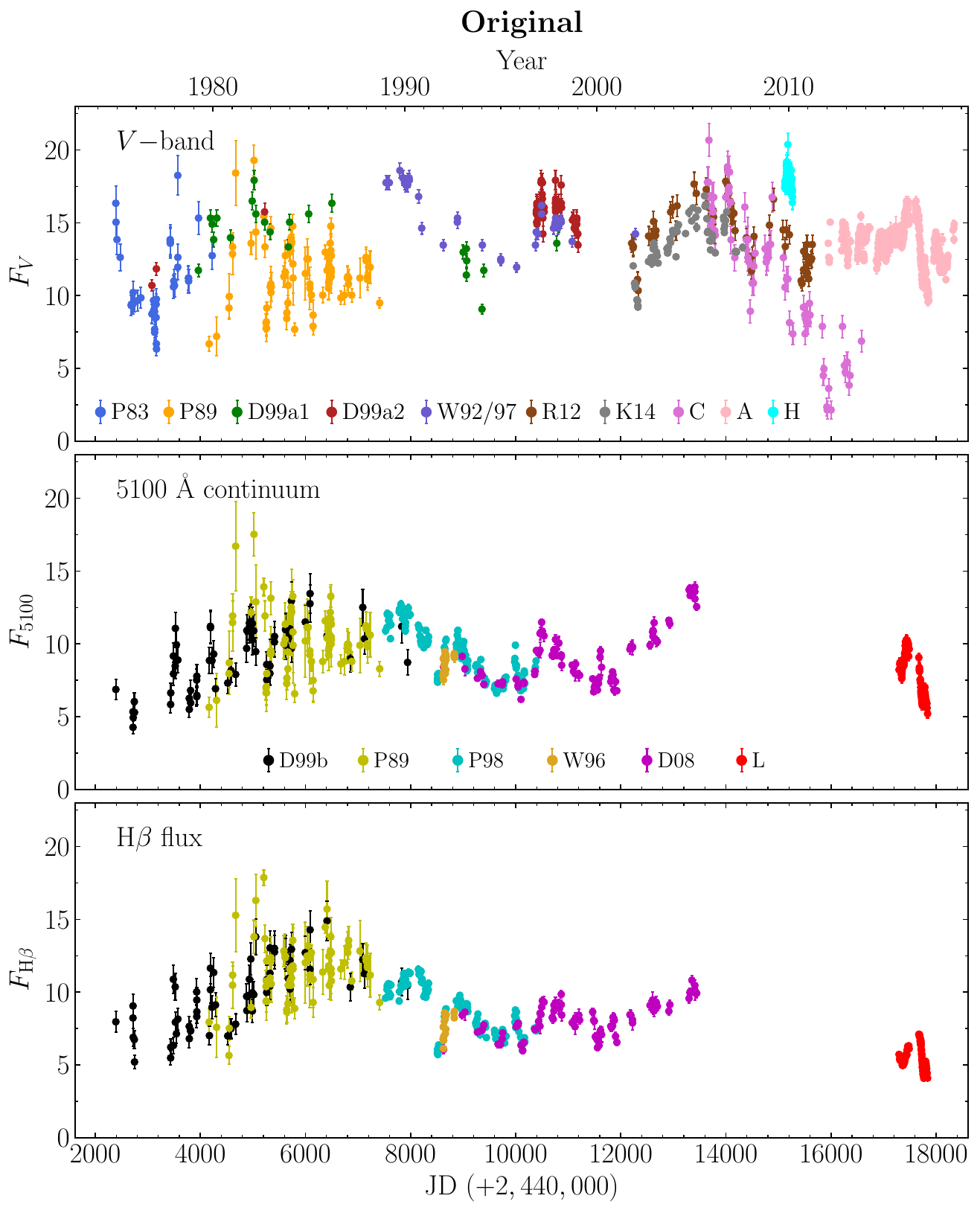}
\includegraphics[width=0.495\textwidth]{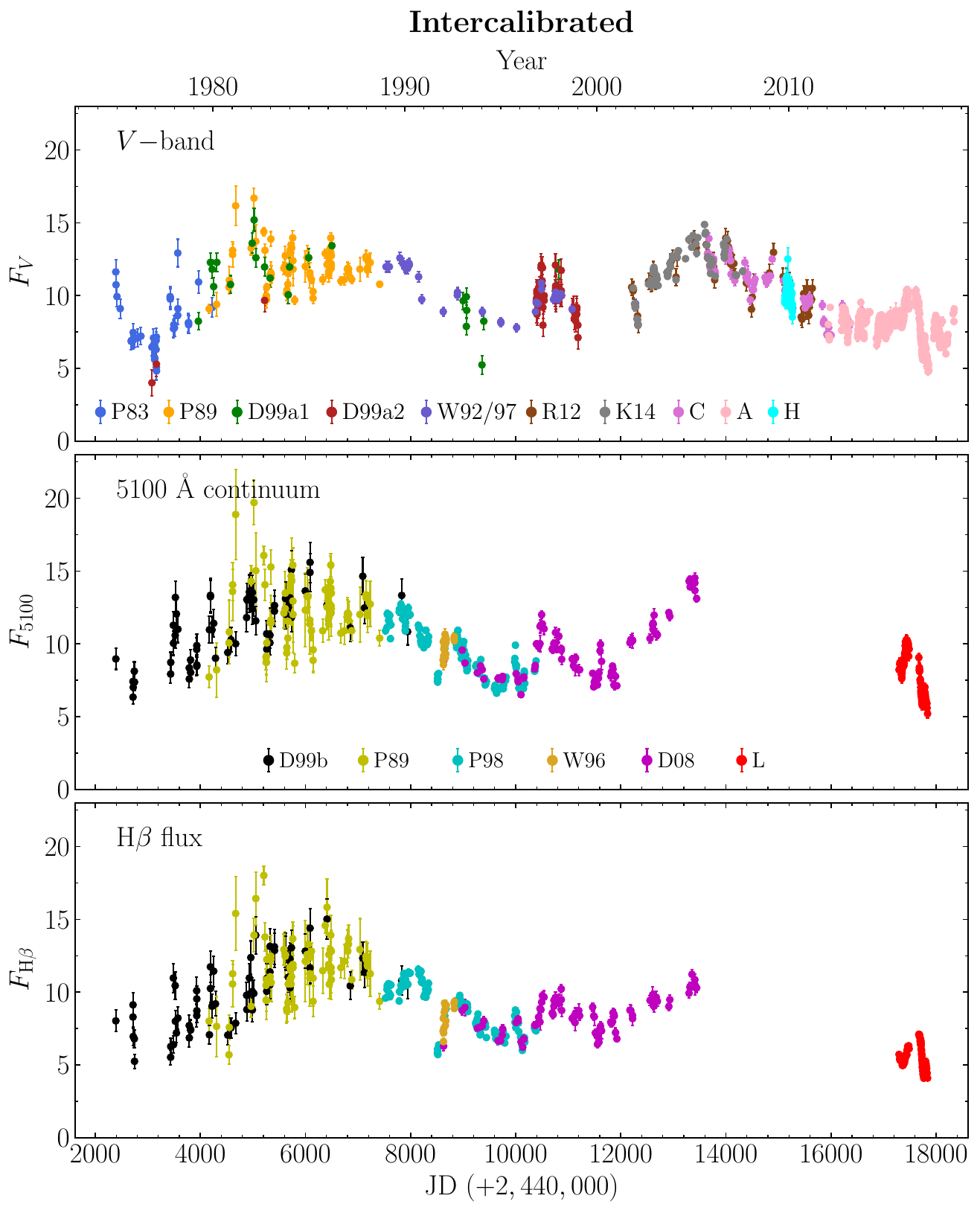}
\caption{(Left) Original and (right) intercalibrated light curves of $V-$band and 5100~{\AA} continuum flux densities 
(in units of $10^{-15}~\rm erg~s^{-1}~cm^{-2}~\text{\AA}^{-1}$), and H$\beta$ integrated fluxes
(in units of $10^{-13}~\rm erg~s^{-1}~cm^{-2}$)
from top to bottom panels. The corresponding references of the labels are listed in Table~\ref{tab_lc}.}
\label{fig_lc}
\end{figure*}

In this paper, we study the variability and analyze the possible periodicity in Ark 120 ($z=0.03271$\footnote{The redshift
is taken from the NASA/Infrared Process and Analysis center (IPAC) Extragalactic Database, corresponding to a luminosity distance of 143 Mpc based on the {\it WMAP-5}
cosmological parameters $H_0=70.5~\rm km~s^{-1}~Mpc^{-1}$, $\Omega_m=0.27$, and $\Omega_\Lambda=0.73$.
}), a nearby radio-quiet AGN with a radio-loudness of $R\approx0.1$ (\citealt{Condon1998, Ho2002}). 
Ark~120 has been intensively observed and investigated in optical/ultraviolet (UV; e.g., \citealt{Kollatschny1981a, 
Kollatschny1981b, Schulz1981, Alloin1988, Marziani1992, Peterson1998,Stanic2000, Popovic2001, Doroshenko2008, Kuehn2008}) and 
X-ray bands (e.g., \citealt{Vaughan2004,Nardini2016, Reeves2016,Gliozzi2017,Lobban2018}).
The published photometric and spectroscopic monitoring data of Ark~120 date back to 1974. Even earlier (1935$-$1950) there %
were historical archival plates collected in the Harvard College Observatory that recorded the photometric images of 
Ark 120 (\citealt{Miller1979}). A recent campaign spectroscopically
monitored Ark 120 for about 100 days in 2017 with the 2.3m telescope of Wyoming Infrared Observatory (\citealt{Du2018}).   
We carried out further observations between 2015 and 2017 %
using the 2.4m telescope of the Yunnan Observatories.
We compile publicly available photometric and spectroscopic data
of Ark~120 and obtain a total temporal baseline stretching over four decades. By visual inspection, the long-term variations in the optical %
continuum display a wave-like pattern with a possible period of $\sim$20 yr. Moreover, the broad H$\beta$ profile of Ark~120 
is highly asymmetric and varies strongly with time (\citealt{Doroshenko2008, Du2018}), a remarkable analogy to what was discovered in NGC~5548 %
(\citealt{Bon2016, Li2016}).

The paper is organized as follows. Section 2 introduces briefly our new monitoring campaign and then presents our construction 
of a long-term database for Ark 120. Section~3 studies the long-term optical variability and measures the power spectral density (PSD).
In Section~4, we perform detailed periodicity analysis for the optical light curve and estimate the associated false alarm probability.
In Section~5, we summarize the long-term variations of the broad H$\beta$ profile. A brief discussion on the possible periodicity 
is given in Section 6 and a conclusion is given in Section 7.

\section{Observations and Historical Database}

\subsection{New Observations between 2015 and 2017}
We carried out a two-year observation campaign between 2015 and 2017 using the Lijiang 2.4m telescope, located 
in Yunnan Province, China, operated by the Yunnan Observatories.
We used the Yunnan Faint Object Spectrograph and Camera (YFOSC) with Grism 14, which covers 
a wavelength range of 3800-7200~{\AA} and provides a resolution of 1.8~{\AA}~pixel$^{-1}$.
This spectrograph is equipped with a slit long enough to allow us to 
simultaneously observe a nearby comparison star (\citealt{Du2014}). We use this comparison star 
as a reference standard to achieve accurate absolute and relative flux calibrations,
whereas the absolute flux of the comparison star is calibrated by 
observations of spectrophotometric standards during nights of good weather conditions.
This flux calibration method is better than the common way based on the flux constancy of
the narrow [\ion{O}{3}] line as described below. 
However, we confirm that the narrow [\ion{O}{3}]~$\lambda5007$ lines of the calibrated spectra 
are constant with a mean of $0.96\times10^{-13}{\rm~erg~s^{-1}~cm^{-2}}$ and a scatter within 2.5\%.
The slit was fixed at a projected width of $2.5^{\prime\prime}$ and the spectra were extracted
using a uniform window of $8.5^{\prime\prime}$.
The raw spectroscopic data are reduced using standard IRAF version 2.16 routines before absolute flux calibration,
which includes bias subtraction, flat-field correction, and wavelength calibration. %
The details for the data reduction and analysis will be presented in a separate paper that describes %
the reverberation mapping analysis (see also \citealt{Du2014}). 
We obtained  100 spectroscopic observations in total, with a typical exposure time of 30 minutes. 
The 5100~{\AA} continuum flux is measured in a band of 20~{\AA} wide around 5100~{\AA} in the rest frame.
To be consistent with the other datasets, the H$\beta$ flux is measured as follows: 
we first subtract the continuum underneath the H$\beta$ line by interpolating between two continuum bands, 
4760-4790~{\AA} and 5000-5020~{\AA}. We then integrate the continuum-subtracted H$\beta$ flux between 4810~{\AA} and 
4910~{\AA}.

The previously published spectroscopic observations were only to 2005, so there is no temporal overlap with  
our observations. As a result, to align the fluxes with the historical data, we need to correct for
the differences in the host galaxy contamination. Based on the two-dimensional decomposition
of the {\it HST} host galaxy image of Ark~120 by \cite{Bentz2009}, we estimate the relative difference %
of the host galaxy contribution to 5100~{\AA} fluxes though the apertures used in our observations and
in \cite{Peterson1998} to be $\Delta F_{\rm gal}=-(0.9\pm0.3)\times10^{-15}~\rm erg~s^{-1}~cm^{-2}~{\text{\AA}^{-1}}$. %

\begin{figure*}[ht!]
\centering
\includegraphics[width=0.95\textwidth]{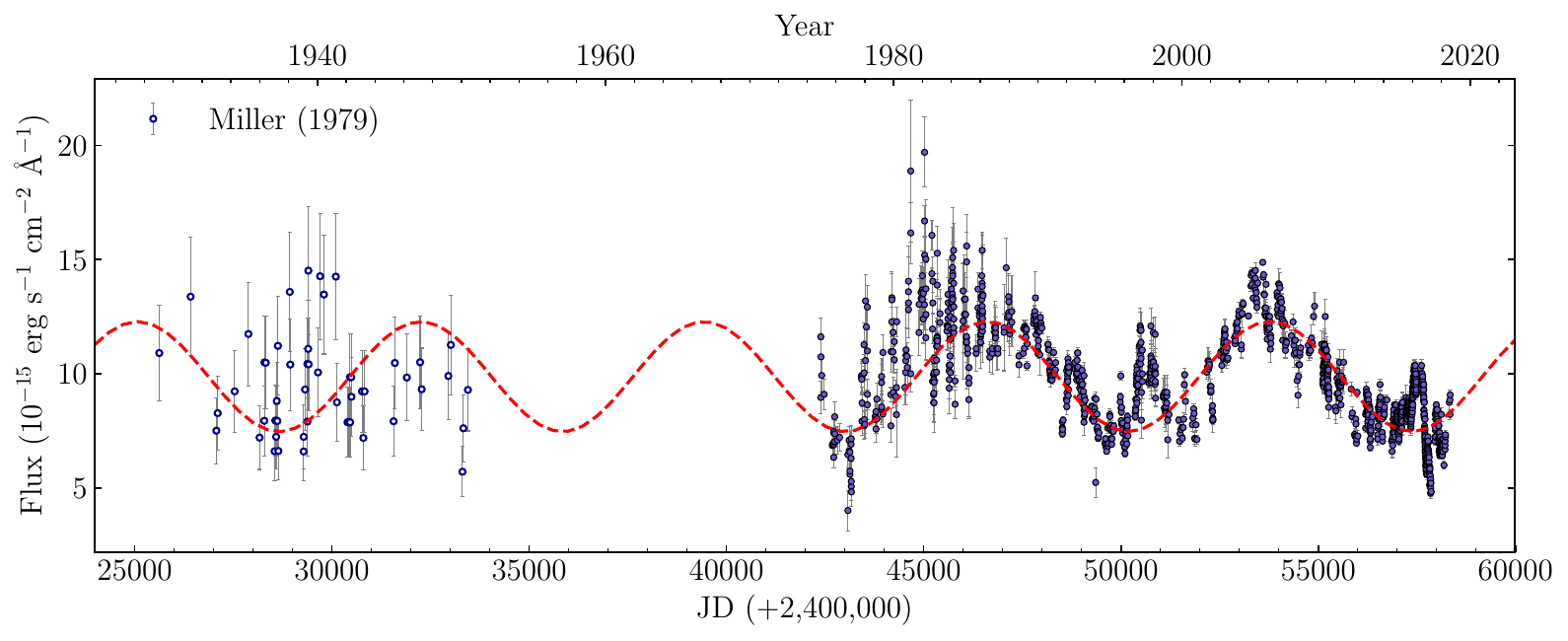}
\caption{Merged light curve of the $V$-band and 5100~{\AA} continuum and the light curve data of \cite{Miller1979} for Ark 120. %
The red dashed line represents a sinusoid with a period of 20.5 yr. Note that it is impossible to %
intercalibrate the dataset of \cite{Miller1979}. We simply adjust the fluxes to make the mean and standard deviation 
to be at the same scales as those of the other datasets (see Section~\ref{sec_miller}). {\it We do not use the dataset of \cite{Miller1979}
for analysis.}}
\label{fig_miller}
\end{figure*}

\subsection{Light Curves of 5100~{\AA} Continuum and H$\beta$ Fluxes}
\label{sec_data}
Table \ref{tab_lc} summarizes the photometric and spectroscopic datasets published to date for 
Ark~120 in the literature. These datasets use various apertures and different flux standards %
of [\ion{O}{3}]~$\lambda5007$ for absolute spectral calibrations.
Intercalibration is needed to correct for these differences.

For spectroscopic data, we select the dataset of \cite{Peterson1998} as 
a reference and scale the other datasets to a common absolute flux %
of [\ion{O}{3}]~$\lambda5007$, $F([\text{\ion{O}{3}}])=0.91\times10^{-13}{\rm~erg~s^{-1}~cm^{-2}}$ as the reference
dataset used%
\footnote{It is possible that the narrow [\ion{O}{3}] line has a long-term secular variation as in NGC~5548 
(\citealt{Peterson2013}). We do not include such a variation in our intercalibration as there were no reliable absolute flux 
measurements over the whole period of the datasets.}.
To account for the different aperture sizes, we apply a scale factor $\varphi$ and a flux modulation 
$G$ to the 5100~{\AA} flux densities and H$\beta$ fluxes of each dataset with respect to the reference 
using (e.g., \citealt{Peterson1995, Li2014}) %
\begin{equation}
F_{\lambda}(5100\text{\AA}) = \varphi F_{\lambda}(5100\text{\AA})_{\rm obs} + G,
\end{equation}
and 
\begin{equation}
F({\rm H\beta}) = \varphi F({\rm H\beta})_{\rm obs}.
\end{equation}
The values of $\varphi$ and $G$ are determined by comparing the closely spaced measurements of the two datasets. 
The interval for comparison is adopted to be 50 days for the datasets 
D08 and W96 in Table~\ref{tab_lc}, which yields a good temporal overlap. 
For the  datasets D99b and P89 there is a short interval (two yr) of temporal %
overlap with the other datasets. Meanwhile, the sampling of these two datasets is poor. 
We thus increase the interval for comparison up to 100 days to obtain a sufficient number of nearly
contemporary observations. Since we concentrate on long-term variations, %
the choice of the interval has no influence on our analysis. 
It is impossible to intercalibrate the H$\beta$ fluxes of our new 
observations with the other datasets, because there is no temporal overlap. We 
present the light curve of the H$\beta$ fluxes for the purpose of illustrating the variation
behaviour of the H$\beta$ line. We do not use it for any subsequent periodicity analysis. %

For $V-$band photometric data, we first convert the magnitudes into flux  densities 
by adopting the zero point $F_\lambda(V=0)=3.92\times10^{-9}\,{\rm erg~s^{-1}~cm^{-2}}\textrm{\AA}^{-1}$ 
(\citealt{Johnson1966}). We then scale the flux densities to align with the 5100 {\AA} flux densities
by again applying a scale factor ($\varphi$) and flux adjustment ($G$)
to bring them to a common flux scale with the light curve of the 5100~{\AA} continuum using 
\begin{equation}
F_{\lambda}(5100~\text{\AA}) = \varphi F_{\lambda}(V)_{\rm obs} + G.
\end{equation}
The interval for comparison is adopted to be 50 days for all the datasets.
Table~\ref{tab_lc} also lists the values of $\varphi$ and $G$ for all the spectroscopic and photometric datasets.
In the right panels of Figure~\ref{fig_lc}, we plot the intercalibrated light curves of $V-$band and 5100~{\AA} flux densities, and H$\beta$
integrated fluxes. The original light curves without intercalibration are plotted in the left panels of Figure~\ref{fig_lc} for 
comparison. Appendix~A tabulates a portion of the merged light curve of the 5100~{\AA} and $V$-band fluxes, while the entire 
light curve data is available in the online journal.

\subsection{The Data Set of \cite{Miller1979}}
\label{sec_miller}
\cite{Miller1979} compiled the $B$-band data of Ark 120 over 1935-1950 using the archival plate collection of the 
Harvard College Observatory. Albeit with poor cadence and low measurement accuracy, this dataset still provides
precious historical variation information of Ark 120. Indeed, \cite{Miller1979} detected significant 
flares around 1937-1939. We digitalize Figure 1 of \cite{Miller1979} and obtain the $B$-band magnitude data.
It is impossible to intercalibrate this dataset with the other datasets described above. Despite this, 
a useful comparison is possible by converting the $B$-band magnitudes into fluxes 
and then simply adjusting the fluxes to make the mean and standard deviation 
to be at the same scales as those of the other datasets. {\it We do not include this dataset for subsequent 
analysis}, but take it as an auxiliary dataset to illustrate the long-term variations of Ark~120.
Figure~\ref{fig_miller} plots the light curve data of \cite{Miller1979} together with the merged light curve of the %
$V-$band and 5100~{\AA} continuum.

\begin{figure}[t!]
\centering
\includegraphics[width=0.48\textwidth]{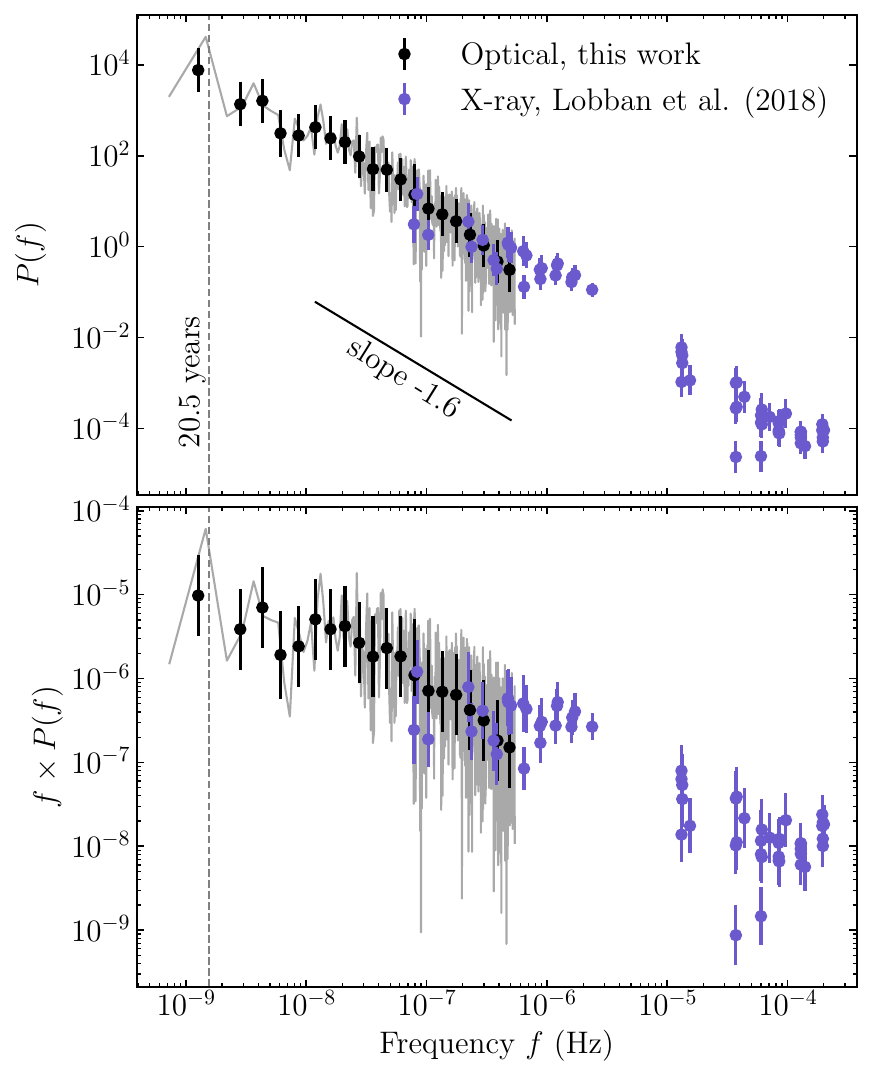}
\caption{(Top) Power spectral density $P(f)$ and (bottom) $f\times P(f)$ of Ark 120 in the optical and X-ray band (0.3-10 keV). 
The normalization of the X-ray PSD is adjusted to match the optical PSD. Blue points represent the 
binned PSD and the grey line represents the raw PSD. Slate blue points represent the X-ray PSD from \cite{Lobban2018}. %
Vertical dashed lines correspond to the period of 20.5 yr.}
\label{fig_psd}
\end{figure}

\subsection{Long-term Optical Variations}
\label{sec_var}
As shown in Figure~\ref{fig_lc}, all of the light curves of the $V-$band, 5100~{\AA} continuum and H$\beta$ integrated 
fluxes oscillate with a wave-like pattern. In the merged light curve plotted in Figure~\ref{fig_miller},
there are clearly two major crests around 1985 and 2005 and two major troughs around 1995 and 2015. %
Meanwhile, one prominent flare appears around the bottom of each major trough. %
Both flares rise rapidly in flux by more than 50\% within 200 days. The latest flare (2016-2018) then undergoes 
an even more sudden drop within the same timescale. 
All the separations between 
peaks/toughs and between the locations of the flares are roughly 20 yr%
\footnote{We note that some numerical simulations of supermassive black hole binary systems produce a similar 
variation pattern (crests/troughs and flares) in mass accretion rates onto the binary black holes (e.g, 
\citealt{Bogdanovic2008, Bogdanovic2009,Miranda2017}).}.
If we extrapolate such a repeated
pattern to the period of \cite{Miller1979}'s dataset, we find that the time (1937-1939) of the significant flare
reported in \cite{Miller1979} remarkably coincides with the anticipated time of recurring flares with a period of 
$\sim20$ yr (see Figure~\ref{fig_miller}). Unfortunately, the data from \cite{Miller1979} are very noisy and scattering, 
therefore, we are unable to be sure whether the variations follow the wave-like pattern seen in the light curve from 1974 to 2018.
On a short timescale of months, Ark~120 also undergoes strong variations, %
which are commonly believed to originate from local instabilities of the accretion disk (e.g., \citealt{Czerny2008, Kelly2009}).

\begin{deluxetable*}{cccccccccc}
\tablewidth{1.0\textwidth}
\tablecolumns{10}
\tablecaption{Inferred values of PSD model parameters.\label{tab_par}}
\tablehead{
\colhead{Model}    &
\colhead{~~~~~~$\log A$~~~~~~}      &
\colhead{~~~~~~$\alpha$}~~~~~~&
\colhead{~~~~~~$\beta$~~~~~~}  &
\colhead{~~~~~~$\log f_{\rm b}$~~~~~~} &
\colhead{~~~~~~$\log A_{\rm p}$~~~~~~} &
\colhead{~~~~~~$\log f_{\rm p}$~~~~~~} &
\colhead{~~~~~~$\log \omega_{\rm p}$~~~~~~} &
\colhead{$\log K$}  &
\colhead{FAP}\\
\colhead{}   &
\colhead{}   &
\colhead{}   &
\colhead{}   &
\colhead{(day$^{-1}$)}   &
\colhead{}  & 
\colhead{(day$^{-1}$)}   &
\colhead{(day$^{-1}$)}   &
\colhead{}  &
\colhead{}
}
\startdata
SPL          &  $-2.2\pm0.2$         & $1.6\pm0.1$ & \nodata     & \nodata       & \nodata      & \nodata        & \nodata      & 0              & $8\times10^{-4}$\\
DRW          &  $\phm{+}4.3\pm0.4$   & \nodata     & \nodata     &  $-3.6\pm0.2$ & \nodata      & \nodata        & \nodata      & $-2.37\pm1.00$ & $5\times10^{-4}$ \\
BPL          &  $\phm{+}3.3\pm1.3$   & $2.0\pm0.6$ & $1.6\pm1.1$ & $-3.3\pm0.8$  & \nodata      & \nodata        & \nodata      & $\phm{-}0.31\pm0.82$ & $1\times10^{-3}$\\
SPL+Periodic &  $-2.1\pm0.3$         & $1.6\pm0.1$ & \nodata     & \nodata       & $-1.2\pm1.3$ & $-3.87\pm0.30$ & $-4.5\pm0.8$ & $-3.34\pm1.14$ &\nodata\\
DRW+Periodic &  $\phm{+}3.4\pm0.5$   & \nodata     & \nodata     & $-3.2\pm0.3$  & $-1.2\pm1.1$ & $-3.88\pm0.23$ & $-4.7\pm0.8$ & $-1.75\pm0.97$ &\nodata \\
BPL+Periodic &  $\phm{+}2.5\pm1.2$   & $2.3\pm0.8$ & $1.7\pm0.9$ & $-2.8\pm0.8$   & $-1.5\pm1.4$ & $-3.80\pm0.36$ & $-4.5\pm0.9$ & $-2.78\pm0.76$ &\nodata 
\enddata
\tablecomments{``SPL'' means the single power-law PSD model, ``DRW'' means the damped random walk model, 
``BPL'' means the bending power-law PSD model, and ``Periodic'' means the periodic PSD component which is parameterized 
to be a Gaussian. $K$ is the Bayes factor given with respect to the SPL model and FAP means the false alarm probability.}
\end{deluxetable*}

\section{PSD Analysis}

In Figure~\ref{fig_psd}, we  show the PSD of the merged light curve. 
We resample the light curve into an even time grid, subtract the mean flux to eliminate the zero-frequency
power,  and then calculate the modulus squared of the discrete Fourier transform at each sampled frequency $f_j$
(\citealt{Uttley2002})
\begin{equation}
|F_N(f_j)|^2 = \left[\sum_{i=1}^{N} x_i\cos(2\pi f_j t_i) \right]^2 + \left[ \sum_{i=1}^{N} x_i\sin(2\pi f_j t_i) \right]^2,
\end{equation}
where $x_i$ is the flux at time $t_i$, $f_j=j/N\Delta T$ with $j=1, ..., N/2$, $\Delta T$ is the sampling interval,
and $N$ is the number of points in the data. The PSD is defined with an adopted normalization as 
\begin{equation}
P(f_j) = \frac{2\Delta T}{N}|F_N(f_j)|^2.
\end{equation}
To compare with the X-ray PSD, we also bin the logarithm of the PSD over a logarithmically spaced frequency grid, with 
a grid width of $\log(1.3)$ and a minimum of two power measurements per bin (\citealt{Uttley2002}). The uncertainties 
of the binned powers are set to be the Poisson noise power. 

\cite{Lobban2018} measured the PSD of Ark 120 in the X-ray band (0.3-10 keV) using $\sim 420$ ks {\it XMM- Newton} monitoring data
and additional {\it Swift}, {\it RXTE}, and {\it NuSTAR} observation data (see their Figure~13). We adjust the normalization 
of the X-ray PSD to align with the optical PSD and superpose it upon the optical PSD in Figure~\ref{fig_psd}. 
Figure~\ref{fig_psd} illustrates that both the optical and X-ray PSDs can be 
universally described by a single power law with a slope of $\sim -1.6$, despite some weak features in the
$f\times P(f)$ plot.  The similar shapes of the 
optical and X-ray PSDs potentially imply that the optical and X-ray variabilities may have the same driving mechanism.
A plausible scenario is reprocessing of X-ray emission into the UV/optical band (\citealt{Guilbert1988}).

We apply three widely used PSD models to fit the optical light curve and compare the relative merits of these three models
using the framework {\texttt{RECON}} developed in \cite{Li2018c}. This framework parameterizes the %
AGN time series in the frequency domain and transforms it back to the time domain to fit the data, and 
infers the model parameters using a Markov-chain Monte Carlo (MCMC) algorithm (see \citealt{Li2018c} for details).
Compared with methods that directly fit PSDs (e.g., \citealt{Vaughan2016}), such an approach can cope with irregularly 
sampled light curves and also naturally take into account the sampling effects, such as spectral leakage and aliasing. 
In addition, the framework calculates the Bayesian evidence and hence the Bayes factor that allows us to %
perform model comparison. The shortcoming of this approach is that the (inverse) Fourier transform and MCMC sampling are %
highly computationally expensive for large numbers of data points.
To save computation time, we bin the light curve every 10 days, reducing the number of points from 1473 to 471.
Such a binning manipulation degrades the maximum frequency that can be obtained from the data. %
Since we mainly concentrate on variations with time scales of years, this manipulation does not influence our main analysis results.
It is worth stressing that in fitting the light curve, we assume independent Gaussian measurement noise. %

Here, the Bayes factor is defined to be the ratio of the posterior probabilities (\citealt{Sivia2006}). For two models, say $M_1$ and $M_2$, %
with equal prior weight, the Bayes factor is simplified to be the ratio of the corresponding Bayesian evidence, %
\begin{equation}
K = \frac{P(M_2|D)}{P(M_1|D} = \frac{P(D|M_2)}{P(D|M_1)},
\end{equation}
where $P(M_{1, 2}|D)$ is the posterior probability; $P(D|M_{1, 2})$ is called the Bayesian evidence, which has 
been marginalized over the whole model parameters; and $D$ represent the observed data. %
A large value of $K>1$ means that model $M_2$ is preferable to model $M_1$. A widely used %
criterion for quantifying the degree of preference is given by \cite{Kass1995}. %

The three PSD adopted models are (1) the single power-law (SPL) model with a form of 
\begin{equation}
P_{\rm SPL}(f) = A f^{-\alpha},
\label{equ_spl}
\end{equation}
(2) the damped random walk (DRW) model with a form of 
\begin{equation}
P_{\rm DRW}(f) = \frac{A}{1 + (f/f_{\rm b})^2},
\label{equ_drw}
\end{equation}
and (3) the bending power-law (BPL) model with a form of
\begin{equation}
P_{\rm BPL}(f) = A \left\{\begin{array}{cc}
          (f/f_{\rm b})^{-\alpha} & {\rm for}~f > f_{\rm b},\\
          (f/f_{\rm b})^{-\beta} & {\rm otherwise}.
         \end{array}\right.
\label{equ_bpl}
\end{equation}
Here $A$, $\alpha$, $\beta$, and $f_{\rm b}$ are free parameters with a restriction of $\alpha > \beta$. 
Throughout the calculations, the prior probabilities 
for $A$ and $f_{\rm b}$ are set to be logarithmic, and the value of $f_{\rm b}$
is limited to the frequency range of the data; the prior probabilities 
for $\alpha$ and $\beta$ are set to be uniform over a range (1, 5) and ($-3$, 5) respectively. 

Table~\ref{tab_par} summarizes the inferred parameter values for the above three PSD models.
The MCMC algorithm adopted in the calculations can not calculate the uncertainty of the obtained
Bayesian evidence in a single running. We ran the algorithm 10 times and set the best %
estimate of the evidence based on the mean and the associated uncertainty based on the standard deviation. 
The Bayes factor of the DRW and BPL models relative to the SPL model are 
$\log K=-2.37\pm1.00$ and $0.31\pm0.82$ respectively.  This indicates that all the three %
PSD models give similarly good fits to the observed data.

\begin{figure*}[t!]
\centering
\includegraphics[width=1.0\textwidth]{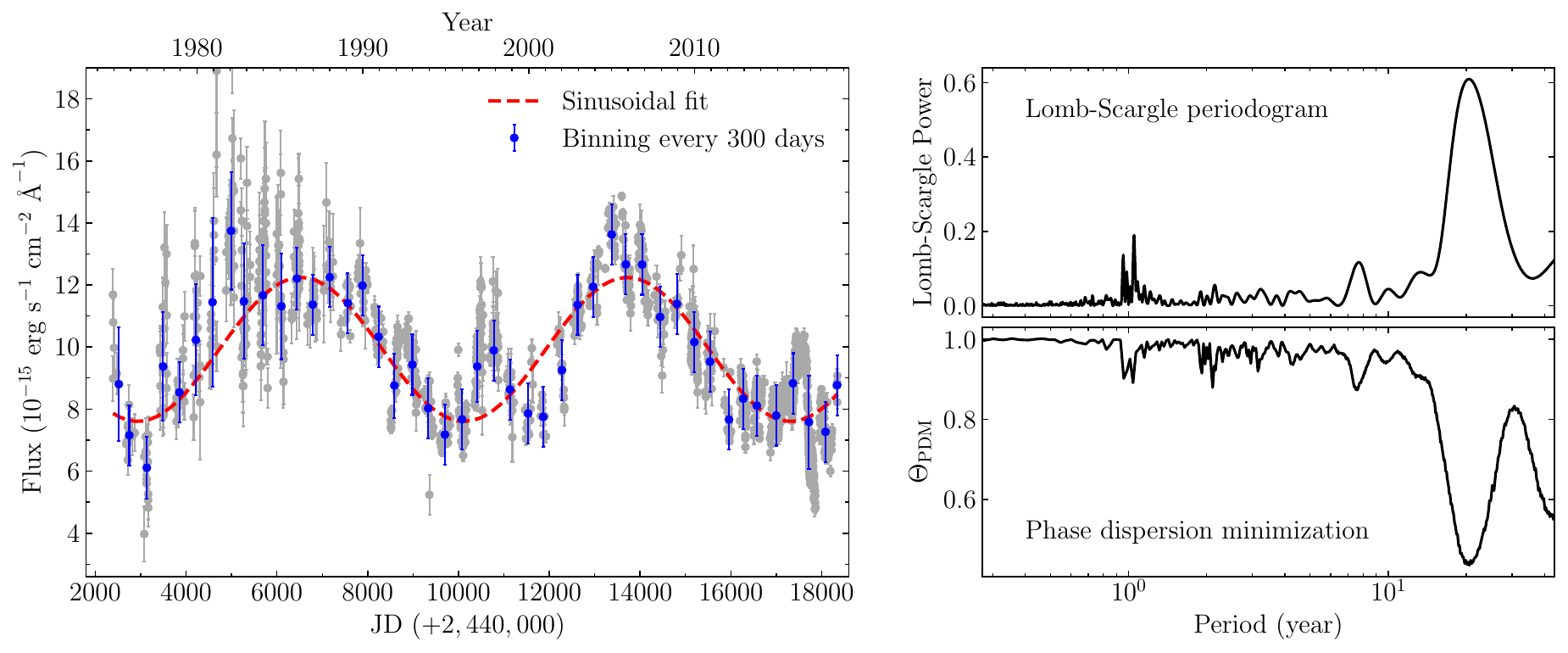}
\caption{(Left) Sinusoidal fit to the binned light curve with a bin width of 300 days (blue points with error bars). Grey points 
with errorbars represent the original (unbinned) light curve.
(Right) The LS periodogram and phase dispersion minimization (PDM) periodogram of the unbinned light curve.}
\label{fig_period}
\end{figure*}
\section{Periodicity Analysis and False-Alarm Probability}
As described above, visual inspection reveals an interval of 20 yr between peaks/troughs, potentially implying that 
a $\sim20$-year periodicity exists in the light curve. Below, we employ several methods to attempt to confirm the periodicity %
and estimate the associated false-alarm probability.

\subsection{Periodicity}
\label{sec_per}
We use three methods to identify possible periodicity in the light curve as follows. %
\begin{itemize}
 \item {\it Lomb-Scargle (LS) periodogram}%
\footnote{We use the module \texttt{LombScargle} in the Python package \texttt{astropy} 
(\citealt{Astropy2018}) to calculate the LS periodogram. %
The argument \texttt{normalization='standard'}
is switched on to implement the ``least-square normalization'' of the periodogram. 
This leads the power of the periodogram to lie between 0 and 1.}.
The top right panel of Figure~\ref{fig_period} shows the LS periodogram of the merged light curve, which peaks at
20.5$\pm$0.1 yr. Here, the uncertainty is determined from the uncertainty of the peak frequency given by
the formula in \citet[see Equation (14) therein]{Horne1986}. 
The LS periodogram essentially assumes a sinusoidal model for the data (\citealt{Scargle1983,VanderPlas2018}).

\item {\it Phase dispersion minimization (PDM) analysis}%
{\footnote{We use the Python module \texttt{CyPDM} to implement 
PDM analysis (\citealt{Li2018a}).}}. The PDM method is widely used to search for non-sinusoidal periodicity by
minimizing the dispersion of the phase-folded data sets (\citealt{Stellingwerf1978}). The bottom right panel of Figure~\ref{fig_period}
plots the PDM periodogram, which is minimized at $20.5$ yr, consistent with the period from the LS
periodogram.

\item {\it Sinusoidal fit}.  
The sampling rate of the light curve is highly irregular and some portions have quite dense cadences. Besides, the light curve  
fluctuates strongly on time scales of weeks/months. To alleviate these two effects, we bin the light curve 
every 300 days, resulting in a total of 45 data points over a baseline of 43 yr. The uncertainties of points
in the newly binned light curve are assigned to be the standard deviations of each bin.
The sinusoidal fitting to the binned light curve yields a period of $19.7\pm0.5$ yr and  sinusoidal amplitude of  
$(2.32\pm0.20)\times10^{-15}\,\rm erg~s^{-1}cm^{-2}\text{\AA}^{-1}$. 
The obtained period slightly depends on the adopted bin width, with the value changing between 19.5 and 20.0 yr for the bin width 
at a range of 100-600 days. 
Throughout the calculations, we use 300 days as the default bin width.
In the left panel of Figure~\ref{fig_period}, we show the binned light curve and the best sinusoidal fit. 
The original, unbinned light curve is also superposed for the sake of illustration.

We note that the LS periodogram is mathematically equivalent to a sinusoidal fit 
(e.g., \citealt{Scargle1983, VanderPlas2018}). Here, we apply the sinusoidal fit to binned light curve data, for the purpose of 
alleviating the influences of the highly irregular sampling rate and strong fluctuations on short time scales.
By contrast, we calculate the LS periodogram using the unbinned light curve data. 
\end{itemize}
In a nutshell, the above three methods generally yield periods of $\sim20$ yr. We use the period $20.5$ yr as the 
fiducial period in the analysis that follows. 

\begin{figure*}[t!]
\centering
\includegraphics[width=0.9\textwidth]{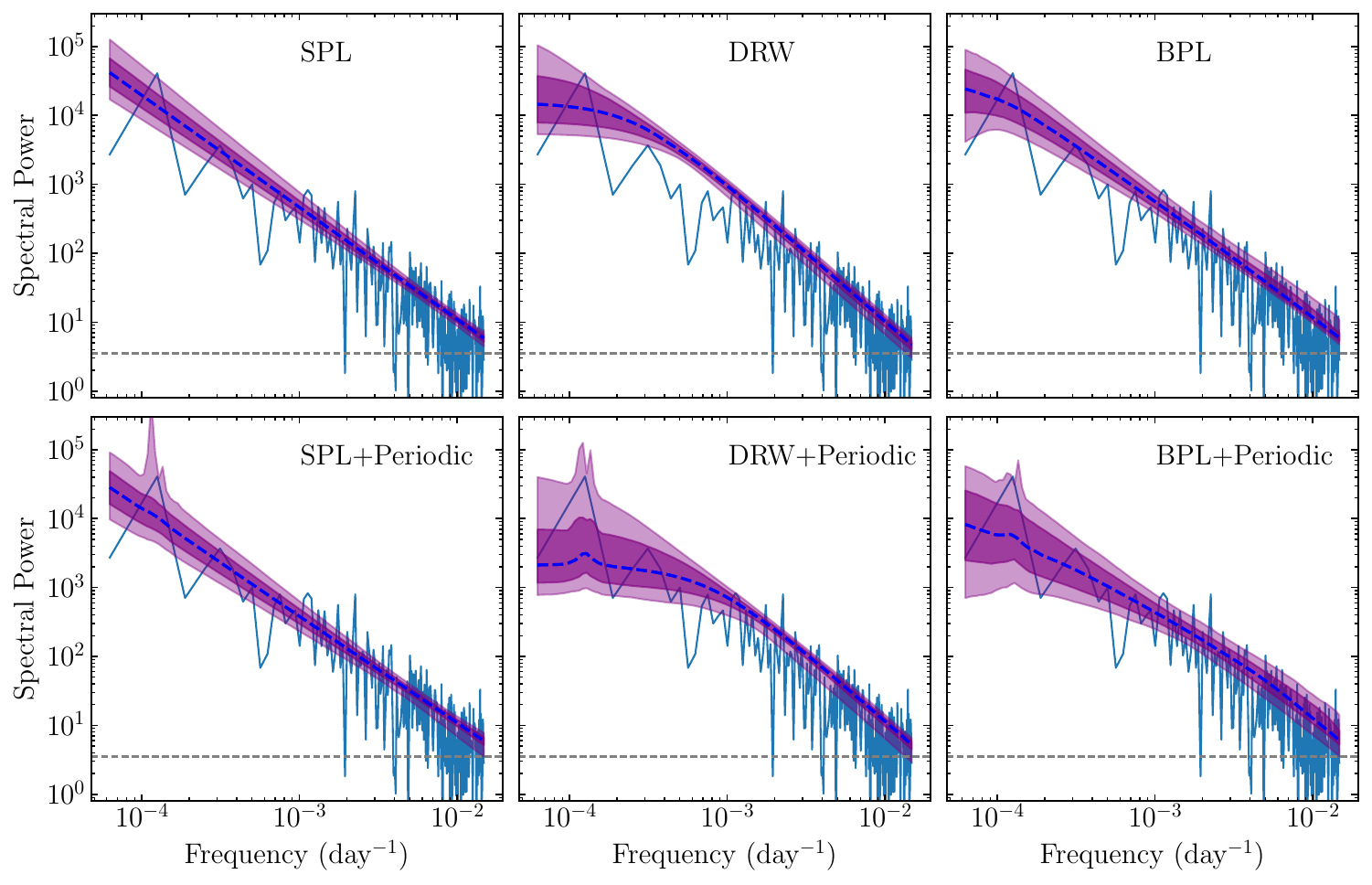}
\caption{The best inferred PSD for different PSD models (blue dashed lines). Shaded areas represent the 1$\sigma$ and 2$\sigma$ error bands. Solid lines represent 
the PSD of the observed data and the grey horizontal dashed line represents the Poisson noise.} %
\label{fig_psd_model}
\end{figure*}

\subsection{Comparison between Periodic and Aperiodic PSD Models}
In Section~\ref{sec_var}, we apply three aperiodic PSD models to fit the light curve of Ark 120. 
We also construct periodic models by including a periodic component into 
the above three PSD models (\citealt{Li2018c, Li2019}). The periodic PSD models have a composite form of
\begin{equation}
P(f) = P_{\rm aperiodic}(f) + \frac{A_{\rm p}}{\sqrt{2\pi} \omega_{\rm p}} \exp\left[-\frac{(f-f_{\rm p})^2}{2\omega_{\rm p}^2} \right],
\end{equation}
where $P_{\rm aperiodic}(f)$ is the aperiodic component that can be $P_{\rm SPL}(f)$, $P_{\rm DRW}(f)$, or $P_{\rm BPL}$ in Equations
(\ref{equ_spl})-(\ref{equ_bpl}), 
and the second term on the right-hand side is the periodic component, which is parameterized to be a Gaussian with %
free parameters $A_{\rm p}$, $f_{\rm p}$, and $\omega_{\rm p}$. The corresponding period of the light curve is $1/f_{\rm b}$.
Since the values of all of these three parameters can span a wide range of magnitudes, their prior probabilities
are set to be logarithmic.
The values of $f_{\rm p}$ are additionally limited by the frequency range of the data.

We tabulate the inferred parameter values for the above periodic PSD models in Table~\ref{tab_par}.
Remarkably, the center of the Gaussian for the three models corresponds to a period of 
$\log(P/{\rm yr})=1.31, 1.32, 1.25$ respectively, with a typical uncertainty of 0.30 dex. %
This is in agreement with the period $\log (P/{\rm year}) = 1.3$ from the LS periodogram.
In Figure~\ref{fig_psd_model}, we show the best recovered PSDs for both aperiodic and periodic models.
The Bayes factors for the three periodic models with respect to the (aperiodic) SPL model are 
$\log K=-3.34\pm1.00$, $-1.75\pm0.97$ and $-2.78\pm0.76$ respectively. 
If adopting $\log K<-2.18$ as the criterion for rejecting a model (\citealt{Kass1995}), 
the obtained Bayes factors imply that for the current observation data, the periodic models are 
not preferable to the aperiodic models. %

\subsection{False-alarm Probability of the Periodicity}
\cite{Vaughan2016} showed that red-noise stochastic processes can easily produce false positive periodicity
in few-cycle light curves of AGNs. The current light curve data of Ark~120 only has roughly two cycles of the period. 
Thus, it is important to appropriately calibrate the false alarm probability (e.g., \citealt{Koen1990,Barth2018, VanderPlas2018}).
False alarm probability essentially measures the probability that an aperiodic variation process (the null hypothesis) can produce 
spurious periodicity as strong or even stronger than that detected in the observed data. 
We perform null hypothesis simulations as follows. For a given aperiodic PSD model, 
we run \texttt{RECON} to obtain a posterior sample of model parameters, from which we randomly draw a subsample to 
generate a set of mock data with exactly the same sampling pattern as the observed data.
To simulate the observation noises, independent Gaussian noise with standard deviation equal to the measurement errors %
are added into the generated artificial data. %

We then apply the three methods described in Section~\ref{sec_per} on the artificial data and analyze the periodicity. %
For sinusoidal fitting, we again bin the light curve every 300 days.
We identify a ``periodic candidate'' if the following 
criteria are satisfied: 
\begin{enumerate}
 \item the LS period of the artificial data is equal to or smaller than that of the observed data, to ensure that the artificial data 
 covers as many cycles of the period as the observed data;
 \item the peak power of the LS periodogram is larger
than or equal to that of the observed data ($P_{\rm LS}\geqslant0.68$) and the minimum of the PDM periodogram is 
smaller than or equal to that of the observed data ($P_{\rm PDM}\leqslant0.41$); 
\item a sinusoid
improves the goodness of fit compared to a constant with $\Delta \chi^2$ greater than or equal to that of 
the binned observed data ($\Delta \chi^2 \geqslant 150$; see Section 4.1);
\item the LS period agrees with 
the periods from PDM analysis and sinusoidal fitting at a level of $\pm 1.0$ yr.
\end{enumerate}
The choice of the second criterion is for the purpose of selecting light curves with strong periodic 
variation amplitudes, the third criterion is to largely ensure roughly equal fluxes of the crests/troughs as seen in 
the light curve of Ark~120, and the fourth criterion aims to reduce possible false detections in the LS periodogram (\citealt{Graham2013}).

The obtained false alarm probabilities are roughly $(5-10)\times10^{-4}$ (equivalent to a significance of $\sim3.3\sigma$) for the above three PSD models.
Combining with the Bayes factors 
calculated in Section 4.2, the values of the false alarm probabilities suggest that the significance 
of the periodicity in Ark 120 is marginal. Continued monitoring of Ark 120 is, therefore, necessary to confirm 
the periodicity.

\begin{figure}[t!]
\centering 
\includegraphics[width=0.47\textwidth]{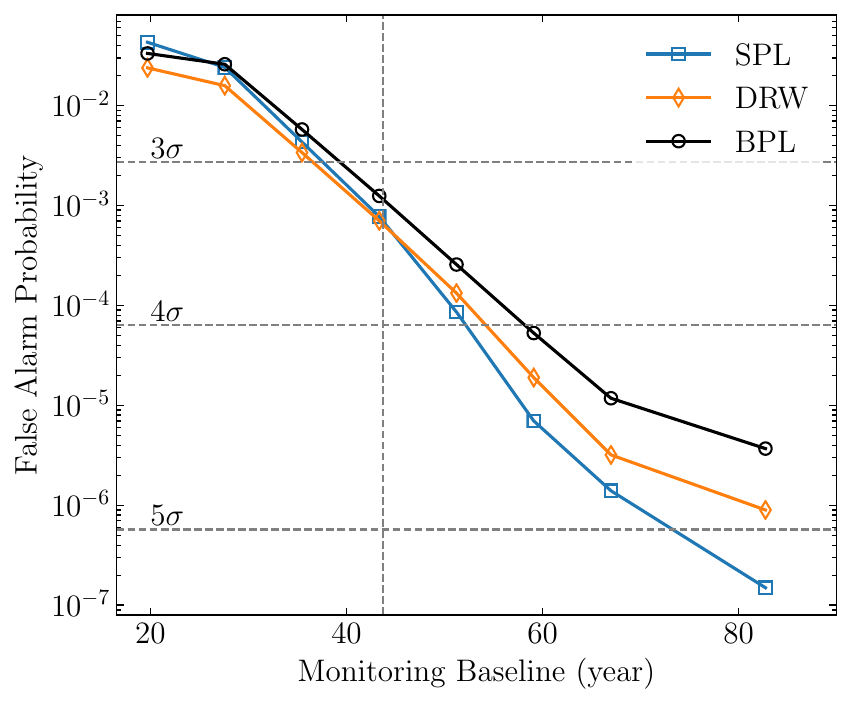}
\caption{Dependence of the estimated false alarm probability on the monitoring time length. The vertical dashed line %
represents the baseline of the observed data. Beyond the data baseline, the false alarm probability is calculated 
using artificial light curves for Ark 120. Note that the false alarm probabilities depend on the %
configurations of artificial data, which are described in Section 4.4.} %
\label{fig_fap}
\end{figure}

\subsection{Requirement of Temporal Baseline for the Periodicity Detection}
The preceding analysis shows that the evidence for the periodicity is inconclusive for the current data. %
A question then arises: how long do we need to monitor Ark 120 to ascertain that the periodicity is significant 
at a given confidence level? To address this question, we make the following assumptions: (1) the long-term variations
are sinusoidal with the parameters given by the sinusoidal fitting in Section 4.1; and (2) the variations at short time scales 
follow a single power-law PSD with the parameters $\log A=-2.2\pm0.2$ and $\alpha=1.6\pm0.1$ from Section~3. The single 
power law is set to flatten to a constant below $1.0\times10^{-3}~{\rm day}^{-1}$ to confine the generated variations to short 
time scales.  We generate a artificial light curve by summing up the above two variation components. %
Following the typical configurations of the observed data, 
the cadence of the artificial light curve is set to be 10 days apart, and the seasonal gap is set to last three months. %
A Gaussian noise with a standard deviation of $0.37\times10^{-15}~\rm erg~s^{-1}~cm^{-2}~\text{\AA}^{-1}$ is added into 
the artificial light curve. One quarter of points are discarded to mimic bad weather or instrumental problems. 
We then attach the artificial light curve to the observed data and use this new artificial data to calculate 
the false alarm probability as in Section 4.3. Appendix B illustrates three examples of artificial light curves for Ark~120. %

In Figure~\ref{fig_fap}, we show the dependence of the estimated false alarm probability on the monitoring time length
for three aperiodic PSD models. The current baseline of the observed database is 43 yr, corresponding to 
a $\sim3.3\sigma$ confidence level (see Section 4.3). Under the observation configurations described above, an 
additional monitoring period of $\sim$30-60 yr (about 1.5-3 cycles) is required to reach a $5\sigma$ confidence level 
(equivalent to a false alarm probability of $6\times10^{-7}$). We stress that the adopted observation configurations are 
quite conservative and, therefore, the obtained monitoring length is just a conservative estimate. Higher data quality
(e.g., higher signal-to-noise ratio and smaller seasonal gap) may reduce the required monitoring length.

\begin{figure}[!t]
\centering
\includegraphics[width=0.48\textwidth]{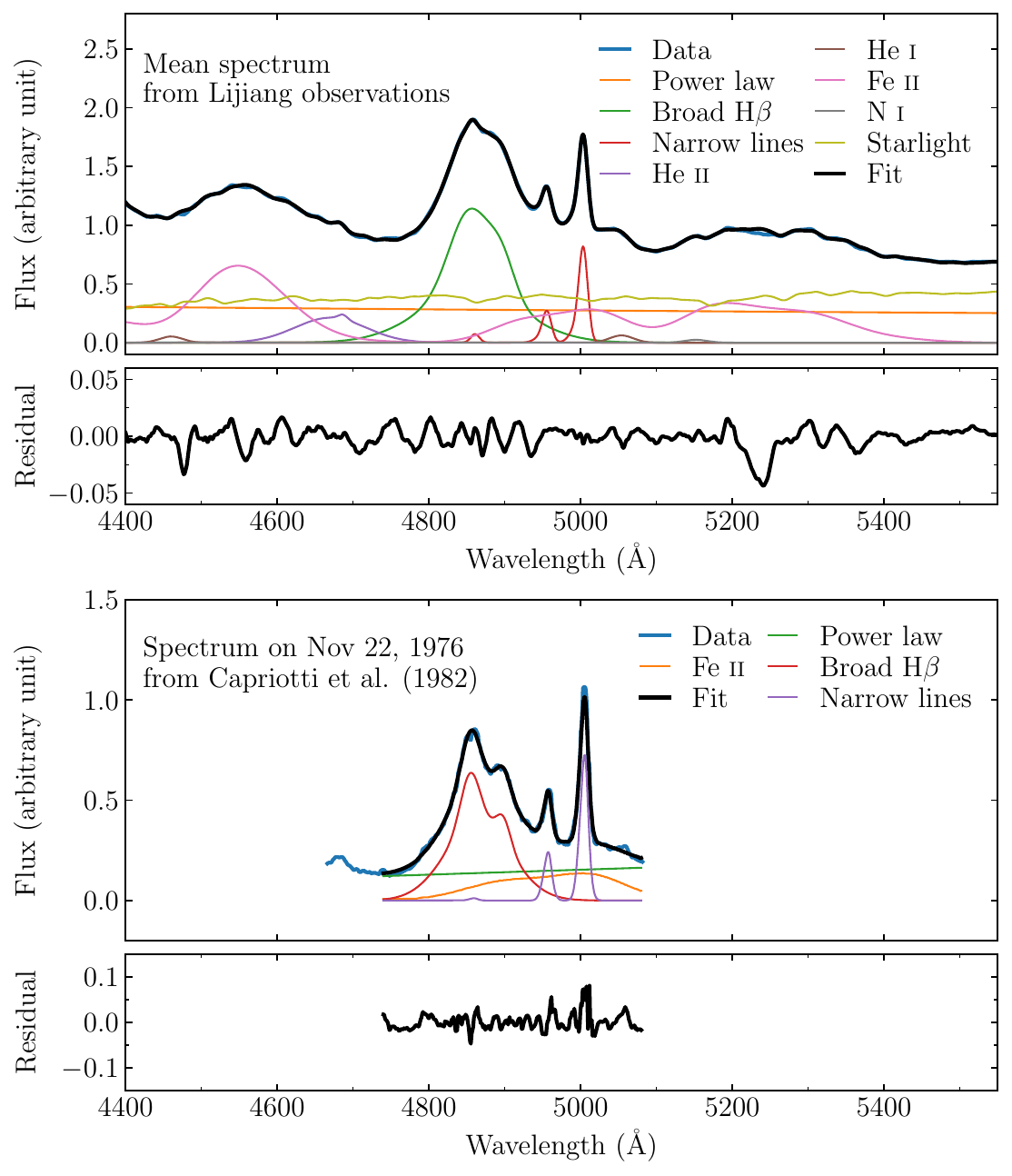}
\caption{Two examples of the extraction of the broad H$\beta$ profile. The top panel shows the mean spectrum of our Lijiang 
observations between 2016 and 2017. The bottom panel shows the historical spectrum from November 22, 1976 digitalized from \cite{Carpriotti1982},
which has narrow wavelength coverage.} 
\label{fig_decomp}
\end{figure}

\begin{figure}[t!]
\centering
\includegraphics[width=0.45\textwidth]{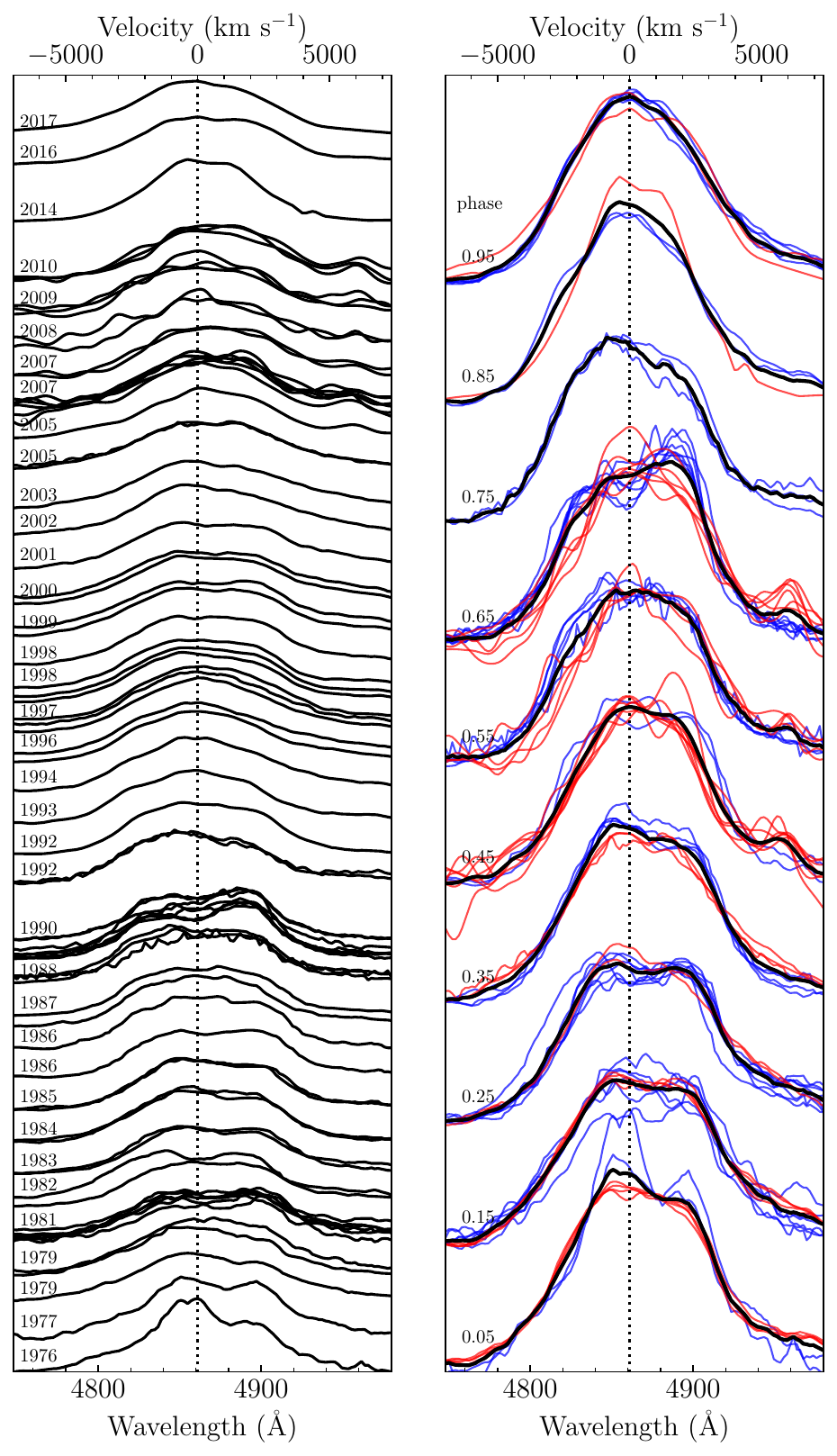}
\caption{(Left) H$\beta$ profiles from 1976 to 2017. (Right) Folded H$\beta$
profile series with phases for a complete cycle using a period of 20.5 yr. Black %
lines represent the overall mean profiles of each phase. Blue and red lines represent
profiles from the first and second cycles of the series respectively. %
Vertical lines show the wavelength center of the H$\beta$ line.} %
\label{fig_prof}
\end{figure}

\section{H$\beta$ Profile Variations}
We also compile the historical H$\beta$ profile of Ark~120. 
The H$\beta$ profiles are mainly from \cite{Stanic2000} and 
\cite{Doroshenko2008}, which presented 10 averaged H$\beta$ profiles over 1977-1990
(see their Figure 3-11) and  21 H$\beta$ profiles over 1992-2005 (see their Figure 5), respectively.
In addition, \cite{Carpriotti1982} displayed six spectra of Ark~120 over 1976-1981 (see their Figure 3), \cite{Korista1992} %
presented 12 spectra over 1981-1989 (see their Figures 1-3), \cite{Peterson1998} show one averaged spectrum from around 1993 (see their Figure 1), 
and \cite{Afanasiev2019} presented one spectrum from 2014 (see their Figure 5). %
We digitalize these figures to obtain the profile data. Since we only concentrate on the 
shape of H$\beta$ profile, the absolute fluxes are no longer important.
Among these references, \cite{Stanic2000}, \cite{Doroshenko2008}, and \cite{Afanasiev2019} had already extracted the 
broad H$\beta$ profiles. We directly use their extracted H$\beta$ profiles. For the other spectra, we isolate %
the broad H$\beta$ profiles following the 
procedure in \cite{Stanic2000}. The narrow H$\beta$ line and \ion{Fe}{2} emission are 
subtracted using a simple spectral decomposition (see \citealt{Stanic2000} for details). 
 The bottom panel of Figure~\ref{fig_decomp} shows an example for decomposing the historical spectrum from %
November 22, 1976 digitalized from \cite{Carpriotti1982}.

\begin{figure*}[t!]
\centering
\includegraphics[width=0.95\textwidth]{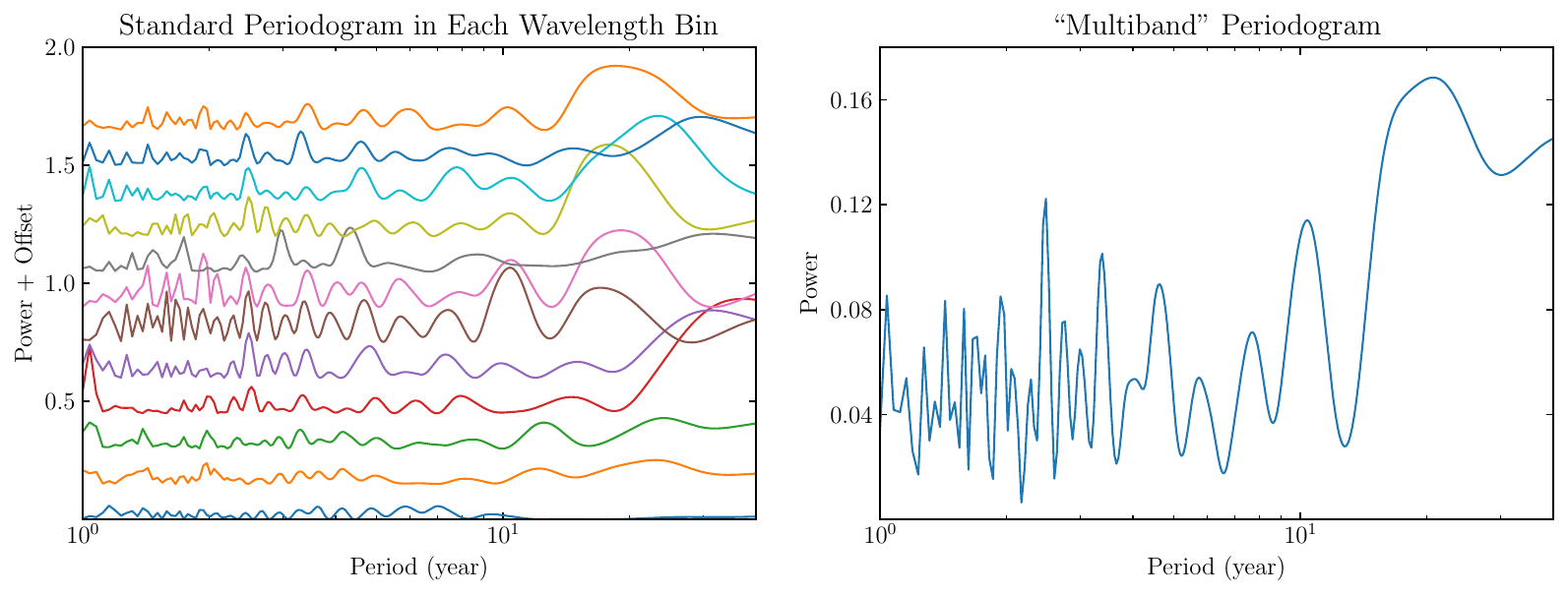}
\caption{(Left) The LS periodogram in 12 wavelength bins of the H$\beta$ profile. (Right) The ``multiband'' LS periodogram that combines 
the information from all the wavelength bins, which is calculated using the Python package \texttt{gatspy} (\citealt{VanderPlas2015s, VanderPlas2015}).}
\label{fig_prof_period}
\end{figure*}

\begin{figure}[t!]
\centering
\includegraphics[width=0.48\textwidth]{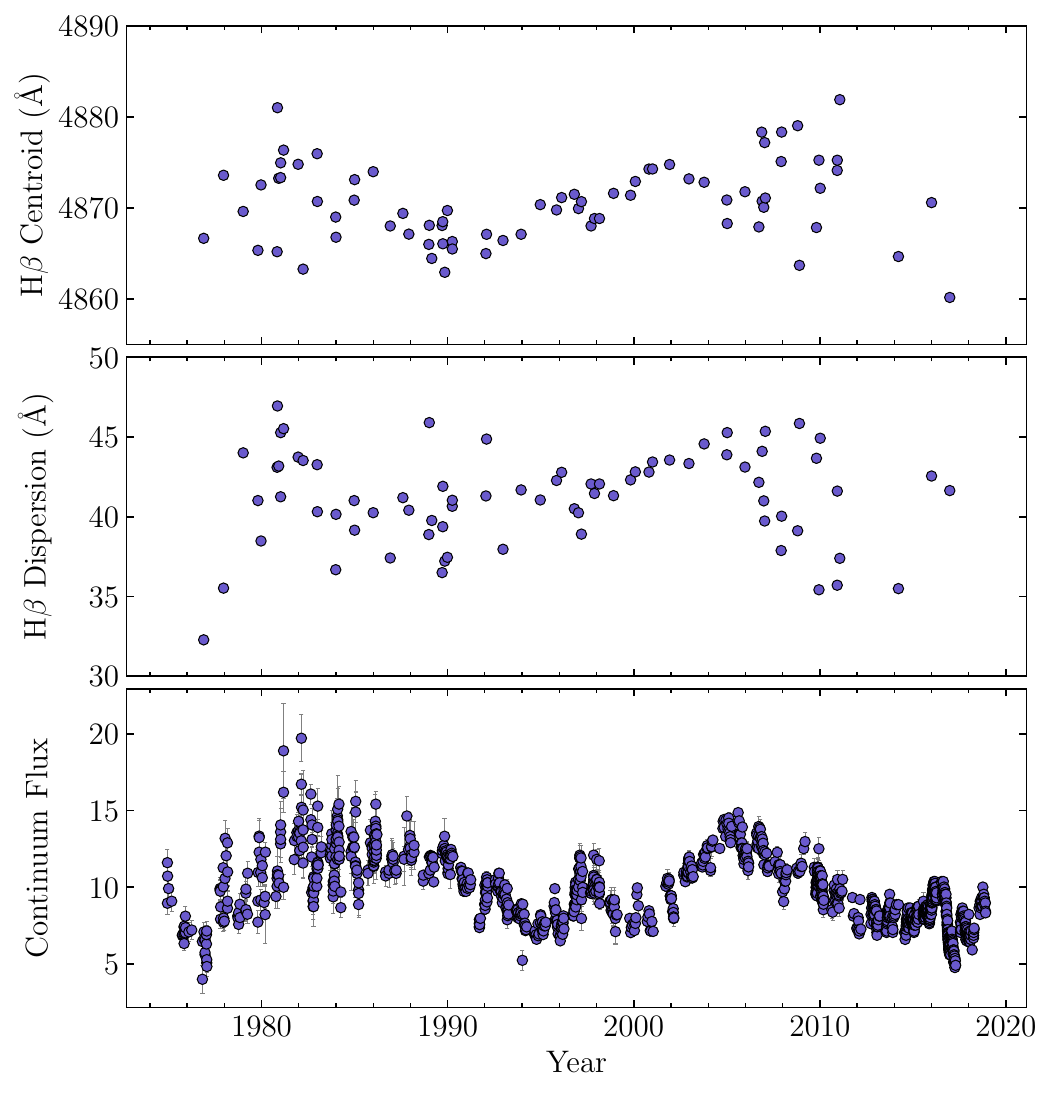}
\caption{(Top) H$\beta$ line centroid, (middle) H$\beta$ line dispersion, and (bottom) continuum fluxes of Ark 120.
The unit of the continuum flux densities is $10^{-15}~\rm erg~s^{-1}~cm^{-2}~\text{\AA}^{-1}$.
The H$\beta$ line centroid and dispersion are calculated over a wavelength range (4741-4981)~\AA.}
\label{fig_hb}
\end{figure}

\cite{Haas2011} displayed one spectrum from 2009 and \cite{Marziani1992} presented four spectra from between 1989 and 1990, 
with both these studies providing us with their electronic data. We queried the NASA/IPAC Extragalactic Database and found 
one spectrum from the 6dF galaxy survey (\citealt{Jones2009}) and one spectrum from the updated Zwicky catalog (\citealt{Falco1999}).
We supply two mean H$\beta$ profiles from our two-year monitoring.  These spectra are of sufficiently good quality to
perform a detailed spectral decomposition. The top panel shows an example of spectral decomposition for the mean 
spectrum between 2016 and 2017 from our Lijiang observations. The decomposition includes a power law for the continuum,
two Gaussians for the narrow [\ion{O}{3}] doublet and H$\beta$ line, 
three Gaussians for the broad emission lines (H$\beta$ and H$\gamma$), an \ion{Fe}{2} model shelf (\citealt{Kovacevi2010}),
and a host galaxy starlight model based on the ELODIE library (\citealt{Prugniel2001,Prugniel2007a, Prugniel2007b}).
Some other emission lines, such as \ion{He}{2}, \ion{He}{1}, \ion{N}{1} etc., are also included when necessary (modelled by one or two 
Gaussians). The flux ratio between [\ion{O}{3}]~$\lambda$5007 and $\lambda$4959 is fixed to be 3. 
A Legendre polynomial of the first order is multiplied to the sum of the above components to account for the errors in the Galactic extinction, flux
calibration, or any other cause that affects the shape of the continuum.
The fitting is implemented using the ULySS code\footnote{The ULySS code is available at \url{http://ulyss.univ-lyon1.fr}.} (\citealt{Koleva2009}). More details of the spectral decomposition were described in 
\cite{Bon2014, Bon2016}.

Furthermore, \cite{Denissyuk2015} collected spectroscopic monitoring data of Ark~120 over %
1976-2013 using a 70-cm telescope. We use 16 of their spectra between 2005 and 2011. %
The spectra are relatively noisy, and we decompose 
the broad H$\beta$ line simply by subtracting the underlying continuum linearly interpolated between 4748~{\AA} and 5020~{\AA}.  
The narrow H$\beta$ and [\ion{O}{3}] lines are not subtracted.

In total, we obtain 76 profiles that cover a time span of 40.1 yr from 1976 to 2017
and show them in Figure~\ref{fig_prof}. We bear in mind that our compilation is incomplete and does not include those studies that investigated
spectral variations of Ark 120 but did not display the spectra. %
The profiles are highly asymmetric and have double peaks (e.g., around 1981 and 1989),
which seem to shift apart and merge again in some epochs.
Moreover, the amplitudes of the red and blue peaks vary independently.
These strong variations of H$\beta$ profiles potentially reflect that the region emitting the broad 
H$\beta$ line (the so called broad-line region) undergoes notable changes. %
Such behavior had also been found in the H$\beta$ profiles of NGC~5548 (\citealt{Bon2016, Li2016}). 
To determine whether there exist periodic variations in the H$\beta$ profiles, we first normalize the H$\beta$ profiles to a uniform integrated flux.
We divide the H$\beta$ wavelength region between 4741 and 4981 {\AA} 
into 12 bins and obtain the ``flux'' in each wavelength bin. We then calculate the ``multiband'' LS periodogram that combines 
the variation information from all the bins using the algorithm developed by \cite{VanderPlas2015}. The right panel of Figure~\ref{fig_prof_period}
plots the ``multiband'' LS periodogram, which peaks around 20 yr. This is remarkably consistent with the period found in the continuum light curve. 
For the sake of comparison, the left panel of 
Figure~\ref{fig_prof_period} shows the standard LS periodogram in each wavelength bin. 
The periodograms show diverse patterns among wavelength bins, meaning that it is impossible to detect 
a period in individual bins under the current data quality.  %

We fold the profiles into 10 uniformly spaced phases using a period of 20.5 yr in the 
right panel of Figure~\ref{fig_prof}. The phase bin width is $\sim2$ yr and each phase bin 
has on average six profiles. It is more evident that in the folded profile series the red and blue peaks 
shift with phases significantly, and their amplitudes vary independently.
For the sake of comparison, we plot the H$\beta$ profiles from the first cycle 
(1976-1996) and second cycle (1996-2017) in blue and red respectively, %
and the overall mean profile in black. The profiles from the two cycles are generally matched in terms of
the separation and amplitude of the double peaks, further supporting that H$\beta$ profiles may change with 
a similar period as the continuum and H$\beta$ fluxes. 

We calculate the H$\beta$ line centroid and dispersion over a wavelength range (4741-4981)~{\AA} from the above spectra 
and show their changes with time in Figure~\ref{fig_hb}. Despite large scattering, both the 
H$\beta$ line centroid and dispersion generally exhibit a wave-like pattern as in the continuum variations. %
It seems that the centroid and line dispersion increase with the continuum flux density. This is difficult to explain 
if assuming that the broad-line region is virialized and its size obeys the tight scaling correlation between sizes %
of broad-line regions and AGN luminosities found in reverberation mapping observations (e.g. \citealt{Kaspi2000, Bentz2013, Du2016}). 
According to this scaling correlation, the size of the broad-line region expands as the AGN luminosity increases, leading to
the line width decreasing if the broad-line region is virialized. However, we stress that the present spectra data are compiled from %
various observations with diverse spectral resolutions, wavelength calibrations, and signal-to-noise ratios. Some of the digitalized spectra 
only cover the wavelength region of H$\beta$ line so that it is impossible to %
recalibrate all the spectra in a self-consistent way. Therefore, more high-quality spectroscopic observations
are needed to test the results in this section. %
 In Appendix A, we present a portion of the online table for 
the H$\beta$ line centroid and dispersion data. Again, because it 
is difficult to reliably estimate the measurement noise for the digitalized spectra, we do not report %
the uncertainties for the calculated H$\beta$ line centroids and dispersions.

\section{ Discussions on the Periodicity}
Regarding the periodicity in AGN variations, there are indeed a variety of theoretic models/interpretations,
including the SMBH scenario (see the discussions
in \citealt{Bon2016}, \citealt{Lu2016}, and \citealt{Li2016}). Some of the interpretations can be directly excluded in Ark~120.
The precessing jet model is implausible as Ark~120 is radio-quiet (\citealt{Condon1998, Ho2002}) so that 
the optical continuum is most likely dominated by the disk emission.
The other interpretations may be tested with the aid of 
high-quality reverberation mapping observations of broad emission lines (e.g., \citealt{Shen2010, Wang2018}). 
In particular, for the SMBH binary scenario, there are two factors that lead to distinct 
reverberation mapping signatures: first, the geometry and dynamics of the broad-line region(s) are different compared to those %
surrounding a single black hole, considering the complicated gravitational potential jointly governed by both black holes
(e.g., \citealt{Sepinsky2007, Popovic2012}); second, if both black holes are active, there are doublet ionizing sources that illuminate 
the broad-line region(s), giving rise to distinguishable patterns in velocity-delay maps of broad emission lines (\citealt{Wang2018}).
Interestingly, a recent reverberation mapping campaign by \cite{Du2018} found a complicated velocity-delay map 
for the H$\beta$ broad-line region of Ark 120, which shows a general decreasing trend from the 
blue ($-3000~{\rm km~s^{-1}}$) to the red ($4000~{\rm km~s^{-1}}$) wing but with a local peak around 
$1000$-$2000~{\rm km~s^{-1}}$. A detailed investigation is needed to uncover the geometry and dynamics 
underlying such a velocity-delay map.

Ark 120 has a stellar velocity dispersion of $\sigma_\star=192\pm8~\rm km~s^{-1}$ (\citealt{Woo2013}), 
resulting in a black hole mass estimate of $M_\bullet=(2.6\pm0.2)\times10^8M_\odot$ from the $M_\bullet-\sigma_\star$
relation for classical bulges (\citealt{Kormendy2013}). By assuming that an SMBH binary resides at the center of Ark~120, the semi-major
axis of the binary's orbit is $a_\bullet = 27.0$ light-days (0.02 pc or 33 $\mu$as) if using 20.5 yr (in the observed frame)
as the orbital period. Such a separation can be potentially spatially resolved either by the %
Event Horizon Telescope with an angular resolution of $\sim20~\mu$as (\citealt{Fish2016}) if each black hole produces radio emission,  
or by the {\it Gaia} satellite with an angular resolution down to $9~\mu$as (\citealt{Dorazio2018}). %
On the other hand, the characteristic strain amplitude of the gravitational wave emission is of the order of $h_s\sim10^{-17}$, %
which is in the sensitivity range of next-generation pulsar timing arrays (\citealt{Janssen2015}). %
The previous H$\beta$ reverberation mapping observations of Ark 120 show a typical size of 30-40 light-days for the H$\beta$ 
broad-line region (\citealt{Peterson1998, Du2018}), indicating that the binary orbit is 
slightly smaller than or comparable with the H$\beta$ broad-line region. In such cases, the interaction between the binary black 
holes and the broad-line regions is therefore important to shape the broad-line profiles.

\section{Conclusion}

We compile the historical archival photometric and spectroscopic data of Ark~120 over four decades.
The long-term variations of both the optical continuum and H$\beta$ integrated 
fluxes exhibit a wave-like pattern. We analyze the periodicity using various methods and generally 
obtain a period of $\sim20$ yr. The estimated false alarm probability is about $1\times10^{-3}$.
The comparison between aperiodic and periodic models based on the Bayes factors suggests 
that the periodicity is inconclusive using the current data. Continued monitoring of Ark~120 is needed to track 
more cycles to eliminate the false positive rate and confirm the periodicity.
The broad H$\beta$ line shows double peaks, which vary strongly with time. 
Using the ``multiband'' LS periodogram developed by \cite{VanderPlas2015}, 
we find that the overall H$\beta$ profile also varies with the same suggested period in the continuum.
These observations make Ark~120 to be one of the nearest %
AGNs with possible periodic variability, a remarkable analogy to NGC 5548 (\citealt{Li2016, Bon2016}).

Although the evidence for periodicity is inconclusive using the current data, %
its abundant spectroscopic monitoring data still make Ark~120 a good laboratory for studying %
the origin of asymmetric, rapidly varying broad emission lines in general and evolution of SMBH binaries, particularly %
if the possible periodic variations stem from the binary's orbital motion.

\acknowledgements{
We thank the referee and the statistics editor for their useful suggestions that improved the manuscript. 
We thank M. Haas for providing the monitoring data of Ark 120 and P. Marziani for providing the spectra of Ark 120.
We acknowledge the support from the staff of the Lijiang 2.4 m telescope.
This research is supported in part from the National Key R\&D Program of China (2016YFA0400701 and 2016YFA0400702),
from the Key Research Program of Frontier Sciences of 
the Chinese Academy of Sciences (CAS; QYZDJ-SSW-SLH007), and from the National Natural Science Foundation of China (NSFC; 
11690024, 11773029, 11833008, 11873048, and U1431228).  Y.R.L. acknowledges financial support from the NSFC (11573026), 
the Strategic Priority Research Program of the CAS (XDB23000000), and 
the Youth Innovation Promotion Association CAS. 
K.X.L. acknowledges support from the Light of West China Program of the CAS (Y7XB016001).
L.C.H. acknowledges financial support from the Kavli Foundation and Peking University.
W.H.B acknowledges financial support from the National Key R\&D Program of China (2017YFA0402703).
L.C.P., E.B., and N.B. acknowledge support from the Ministry of Education and Science of Serbia through 
the projects ``Astrophysical Spectroscopy of Extragalactic Objects'' (176001) and ``Gravitation and structure of universe on large scales'' (176003).
This research has made use of the NASA/IPAC Extragalactic Database (NED), which is operated by the Jet Propulsion Laboratory, 
California Institute of Technology, under contract with the National Aeronautics and Space Administration.
}
\software{ Astropy (\citealt{Astropy2018}), 
CyPDM (\citealt{Li2018a}), gatspy (\citealt{VanderPlas2015s}), 
RECON (\citealt{Li2018b}), ULySS (\citealt{Koleva2009}).}

\appendix
\section{Online Data for the Optical Light Curve and H$\beta$ Line Centroids and Dispersions}
 We tabulate our compiled optical light curve data of Ark 120 between 1974-2018 in Table~\ref{tab_lcsim} and 
the H$\beta$ line centroid and dispersion data in Table~\ref{tab_shift}. Both Tables~\ref{tab_lcsim} and \ref{tab_shift}
show a portion of the data and the entire tables are available in a machine-readable form in the online journal.
 Note that most of the compiled spectra are digitalized from published figures with insufficient wavelength coverage 
to estimate the measurement noise, so we do not report the uncertainties of the calculated H$\beta$ line centroids and %
dispersions.

\begin{deluxetable}{cccc}
\tablewidth{1.0\textwidth}
\tablecolumns{4}
\tablecaption{Compiled optical light curve of Ark 120.\label{tab_lcsim}}
\tablehead{
\colhead{JD}    &
\colhead{~~~~~~Flux~~~~~~}      &
\colhead{~~~~~~Error~~~~~~}     &
\colhead{~~~~~~Dataset~~~~~}    \\
\colhead{(+2,400,000)} & 
\colhead{($10^{-15}~{\rm erg~s^{-1}~cm^{-2}~{\text{\AA}}^{-1}}$)} &
\colhead{($10^{-15}~{\rm erg~s^{-1}~cm^{-2}~{\text{\AA}}^{-1}}$)}
}
\startdata
42392.0000 & 11.68 & 0.84 & P83   \\      
42392.4680 &  8.97 & 0.72 & D99b  \\ 
42396.0000 & 10.78 & 0.77 & P83   \\   
42415.0000 &  9.96 & 0.72 & P83   \\   
42476.0000 &  9.12 & 0.67 & P83   \\   
42685.0000 &  6.85 & 0.59 & P83   \\
... %
\enddata
\tablecomments{The datasets are the same as in Table~\ref{tab_lc}. This table is available in its entirety
in a machine-readable form in the online journal. Only a portion is shown here %
to illustrate its form and content.} %
\end{deluxetable}

\begin{deluxetable}{ccc}
\tablewidth{1.0\textwidth}
\tablecolumns{3}
\tablecaption{The centroid ($\bar\lambda$) and dispersion ($\sigma$) data of the broad H$\beta$ line.\label{tab_shift}}
\tablehead{
\colhead{JD}    &
\colhead{~~~~~~~~~~~~~$\bar\lambda({\rm H}\beta$)~~~~~~~~~~~~~}      &
\colhead{~~~~~~~~~~~~~$\sigma({\rm H}\beta$)~~~~~~~~~~~~~}     \\
\colhead{(+2,400,000)} & 
\colhead{(\AA)} &
\colhead{(\AA)}
}
\startdata
43104.9553 & 4866.66 &  32.27 \\
43494.3051 & 4873.59 &  35.52 \\
43879.0498 & 4869.62 &  44.00 \\
44168.3578 & 4865.34 &  41.01 \\
44228.8177 & 4872.53 &  38.48 \\
44545.9451 & 4865.19 &  43.10 \\
... %
\enddata
\tablecomments{This table is available in its entirety
in a machine-readable form in the online journal. Only a portion is shown here %
 to illustrate its form and content.} %
\end{deluxetable}

\section{Examples of Artificial Light Curves} %
\begin{figure*}[t!]
\centering
\includegraphics[width=0.9\textwidth]{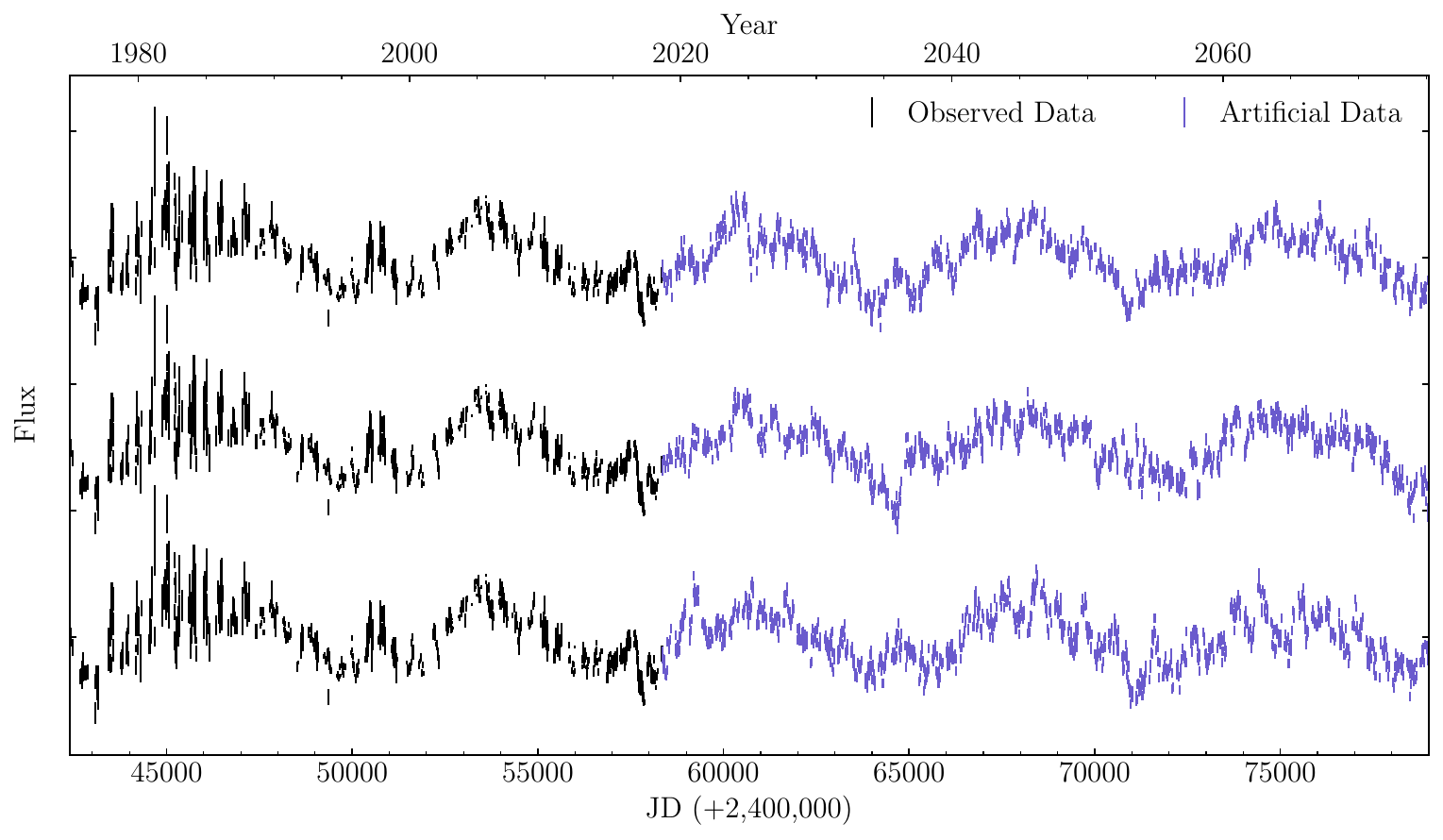}
\caption{Three examples of artificial light curves for Ark 120 used to calculate the false alarm probabilities in Section 4.4. Black points are the observed %
data and blue points are artificial data.} %
\label{fig_artificial}
\end{figure*}

Figure~\ref{fig_artificial} shows three examples of artificial light curves for Ark 120, which are used to determine the %
required baseline to detect the periodicity at a given confidence level. The procedure for generating %
artificial light curves is detailed in Section~4.4. %

\end{document}